\documentclass[12pt]{article}
\usepackage{graphicx,epsfig,cite,graphics}
\usepackage{amssymb}
\usepackage{mathrsfs}
\newcommand{\beq}[1] {\begin{equation}\label{#1} }
\newcommand{\eeq} {\end{equation} }
\newcommand{\bea}[1]{\begin{eqnarray}\label{#1} }
\newcommand{\eea}{\end{eqnarray}}

\def\l{\lambda}
\def\slash {\!\!\!\!/}

\begin{document}

\vspace*{-0.5cm}
\begin{flushright}
OSU-HEP-08-02 \\
HIP-2008-12/TH \\
HRI-RECAPP-08-03\\
NSF-KITP-08-50
\end{flushright}
\vspace{0.2cm}

\begin{center}
{\Large Inverted neutrino mass hierarchy and new signals 
             of a chromophobic charged Higgs at the 
             Large Hadron Collider} 

\vspace*{1.0cm}S. Gabriel$^{(1)\ddagger}$, Biswarup
Mukhopadhyaya$^{(2)\dagger}$, S. Nandi$^{(1)\ast}$ 
and Santosh Kumar Rai$^{(3)\varepsilon}$

\vspace*{0.5cm}
{\sl
$^{(1)}$Department of Physics and Oklahoma Center for High Energy Physics,\\
Oklahoma State University, Stillwater, Oklahoma, 74078\\
$^{(2)}$ Regional Centre for Accelerator-based Particle Physics,\\
Harish Chandra Research Institute, Jhusi, Allahabad 211019, India \\
$^{(3)}$Department of Physics, University
     of Helsinki and Helsinki Institute of Physics,\\ P.O. Box 64,
     FIN-00014 University of Helsinki, Finland\\ \rm}
\end{center}

\vspace*{0.1cm}

\begin{abstract}
We explore the signals of a charged Higgs arising in a two Higgs
doublet model respecting $SU(2)_L \times U(1) \times Z_2$ symmetry
with three singlet right handed neutrinos, $N_R$. The charged Higgs
in this model has negligible coupling with quarks, and has unsuppressed
coupling to leptons and neutrinos. This leads to
novel signatures of the charged Higgs at the LHC, especially
in the case of an inverted neutrino mass hierarchy, in the
form of electrons and muons with missing energy.
\end{abstract}

\vfill

\noindent $^{\dagger}$biswarup@hri.res.in

\noindent $^{\ddagger}$svengab@msn.com

\noindent $^{\ast}$snandi@okstate.edu

\noindent $^{\varepsilon}$santosh.rai@helsinki.fi
\newpage

\section{Introduction}

Although the Higgs boson, the central pillar of the standard electroweak
model (SM) is yet to be observed, there are speculations
on the possibility of a
Higgs sector extending beyond the single Higgs doublet scenario postulated in
the SM. The motivations for such an extended Higgs sector, with masses of the
additional scalars within an experimentally accessible range, are of different
kinds, including the following:

\begin{itemize}
\item Supersymmetry, a widely studied theory for stabilizing
the electroweak scale, which requires at least {\em two} Higgs
doublets \cite{higgshunter}.

\item Little Higgs theories \cite{littlehiggs}, 
which seek to stabilize the electroweak scale
by postulating a low-energy effective theory with several
pseudo-goldstone bosons including the SM-type Higgs.

\item Higgs bosons coming as part of a SU(2) triplet (either as
consequences of a
broken left-right symmetry or introduced in a purely phenomenological
 manner), which helps in the generation of neutrino masses in a type-II
seesaw mechanism \cite{triplethiggs}.
\end{itemize}

The above scenarios all imply the existence of
charged scalar physical states, the experimental signals
of which arise mostly through their coupling
with heavy fermions such as the top and bottom quarks 
and the tau-lepton \cite{chh-searches}. 
In some cases where the charged Higgs are `fermiophobic', interactions with
gauge bosons constitute the search channels \cite{fphobic}. 
However, light fermions
like the electron and the muon are hardly considered relevant
at the primary level, as far as the usually adopted search strategies
for charged Higgs bosons
are concerned. In this note, we suggest the signatures of charged Higgs
in such an unusual channel at the Large Hadron Collider (LHC),
as the consequence of a recently proposed model aimed at explaining
the ultra-small neutrino masses.

The proposal is centred around two Higgs doublets, one of which
($\chi$) couples to all fermions excepting the neutrinos which are
left out by virtue of a $Z_2$ symmetry. The other doublet ($\phi$)
couples only with the charged leptons and the corresponding neutrinos
 which are Dirac fermions in this model \cite{gabriel:2006}. A tiny vacuum
expectation value (vev) of about $10^{-(1 - 2)}$ eV for $\phi$ is
responsible for the smallness of the neutrino masses\footnote
{Though this amounts to fine-tuning, the standard model itself, and
most type-II seesaw models, are finely tuned to at least the same
degree.}. The charged physical scalar which is constituted mostly out
of $\phi$ couples to a charged lepton and a neutrino with large
strength (proportional to the neutrino mass divided by a tiny vev of
the order of the neutrino mass). Importantly, it is `chromophobic' in
nature, in the sense that it has no coupling with quarks.

The available data on  neutrino masses and mixing admit three mass
patterns, namely, normal hierarchy (NH), inverted hierarchy (IH) 
and degenerate neutrinos (DN) \cite{numasharchy}. While the charged
Higgs in this model interacts dominantly with $\tau \nu_{3}$ in the
case of NH (where $m_3 >>m_2 \simeq m_1$), its dominant couplings in
an IH scenario (with $m_2 \simeq m_1 >>m_3$) are with $\mu \nu_{2}$
and  $e \nu_{1}$ in an equitable fashion. As a result, the charged
Higgs scalar, produced, for example, through a Drell-Yan process at
the LHC, will decay into muons and electrons (together with
neutrinos) if one has IH in the neutrino mass patterns. Here we
discuss the detectability of such novel charged Higgs signals. Due
to the striking character of the signals, we mostly discuss the IH
scenario, although we mention the NH case briefly.

We re-iterate the salient features of the model in section 2. Section
three contains a discussion on the proposed signal, the strategies
for eliminating backgrounds, and the predicted numerical results.
We conclude in section 4.

\section{The model and the formalism}

Our proposed model \cite{gabriel:2006} is based on
the symmetry group $SU(3)_c \times SU(2)_L \times U(1) \times Z_2$.
In addition to the usual SM fermions, we have three $SU(2)$ singlet
right-handed neutrinos, $N_{Ri}, i=1-3$, one for each family of
fermions.  The model has two Higgs doublets, $\chi$ and $\phi$.  All
the SM fermions and the Higgs doublet $\chi$, are even under the
discrete symmetry, $Z_2$, while the RH neutrinos and the Higgs
doublet $\phi$ are odd under $Z_2$. Thus all the SM fermions, except
the left-handed neutrinos, couple only to $\chi$. The SM left-handed
neutrinos, together with the right-handed neutrinos, couple only to
the Higgs doublet $\phi$. The gauge symmetry $SU(2) \times U(1)$ is
broken spontaneously at the electroweak 
scale by the vev of $\chi$, while the
discrete symmetry $Z_2$ is broken by a vev of $\phi$, and we take
$\langle\phi\rangle \sim 10^{-2}~eV$. Thus, in our model, the origin
of the neutrino masses is due to the spontaneous breaking of the
discrete symmetry $Z_2$. The neutrinos are  massless in the limit of
exact $Z_2$ symmetry. Through their Yukawa interactions with the
Higgs field $\phi$, the neutrinos acquire masses much smaller than
those of the quarks and charged leptons due to the tiny vev of
$\phi$.

The Yukawa interactions of the Higgs fields with the leptons are
\bea{fermion}
 L_Y =y_{l}\overline{\Psi}^{l}_{L}l_{R}\chi+y_{\nu_{l}}
\overline{\Psi}^{l}_{L}N_{R}\widetilde{\phi}+h.c.,
 \eea
where $\overline{\Psi}^{l}_{L} =
(\overline{\nu}_{l},~\overline{l})_L$ is the usual lepton doublet
and $l_R$ is the charged lepton singlet.  The first term gives rise
to the mass of the charged leptons, while the second term gives a
tiny neutrino mass.  The interactions with the quarks are the same
as in the Standard Model with $\chi$ playing the role of the SM
Higgs doublet. Note that in our model, a SM left-handed neutrino,
$\nu_L$ combines with a right handed neutrino, $N_R$, to make a
massive Dirac neutrino with a mass $\sim 10^{-2}$ eV, the scale of
$Z_2$ symmetry breaking.

 The most general Higgs potential consistent with the $SM \times
Z_2$ symmetry is
\bea{potential} \ V =
-\mu^2_1~\chi^{\dag}\chi-\mu^2_2~\phi^{\dag}\phi+\lambda_1(\chi^{\dag}\chi)^{2}+\lambda_2(\phi^{\dag}\phi)^{2}+\lambda_3(\chi^{\dag}\chi)(\phi^{\dag}\phi)-\lambda_4|\chi^{\dag}\phi|^{2}\nonumber\\-\frac{1}{2}\lambda_5[(\chi^{\dag}\phi)^{2}+(\phi^{\dag}\chi)^{2}].
\eea The physical Higgs fields are a charged field $H$, two neutral
scalar fields $h$ and $\sigma$, and a neutral pseudoscalar field
$\rho$.  In the unitary gauge, the two doublets can be written as

\bea{chi} \chi = \frac{1}{\sqrt{2}}\left(
                                               \begin{array}{c}
                                                 \sqrt{2} (V_\phi/V)H^{+} \\
                                                 h_0 + i (V_\phi/V)\rho
                                                 +V_\chi\\
                                               \end{array}
                                             \right),\nonumber
                                             \eea
\bea{phi} \phi = \frac{1}{\sqrt{2}}\left(
                                               \begin{array}{c}
                                                 -\sqrt{2} (V_\chi/V)H^{+} \\
                                                 \sigma_0 - i (V_\chi/V)\rho
                                                 +V_\phi\\
                                               \end{array}
                                             \right),
 \eea
 where $V_\chi = \langle\chi\rangle$, $V_\phi = \langle\phi\rangle$,
 and $V^{2} = V^{2}_\chi + V^{2}_\phi$.  The particle masses are

 \bea{masses} m^2_{W} = \frac{1}{4}g^{2}V^{2},~ m^2_{H^\pm} 
= \frac{1}{2}(\lambda_4 +\lambda_5)V^{2},~
 m^{2}_\rho = \lambda_5 V^{2},\nonumber
\eea
 \bea{more}m^{2}_{h,\sigma} = (\lambda_1
 V^{2}_\chi +\lambda_2 V^{2}_\phi)\pm \sqrt{(\lambda_1
 V^{2}_\chi -\lambda_2 V^{2}_\phi)^{2}
 +(\lambda_3-\lambda_4-\lambda_5)^{2} V^{2}_\chi V^{2}_\phi}.
 \eea
An immediate consequence of the scenario under consideration is a
very light scalar $\sigma$ with mass
\bea{light} m^{2}_\sigma = 2\lambda_2
V^{2}_\phi[1+O(V_\phi/V_\chi)]. \eea The mass eigenstates $h,
\sigma$ are related to the weak eigenstates $h_0, \sigma_0$ by
\bea{states}
 h_0 = ch+s\sigma,~\sigma_0 = -sh+c\sigma,
  \eea
where c and s denotes the cosine and sine of the mixing angles, and
are given by
\bea{co}
 c = 1+O(V^{2}_\phi/V^{2}_\chi),\nonumber
 \eea \bea{si} s =
-\frac{\lambda_3-\lambda_4-\lambda_5}{2\lambda_1}(V_\phi/V_\chi)+O(V^{2}_\phi/V^
{2}_\chi).
\eea Since $V_{\phi} \sim 10^{-2}$ eV and $V_{\chi} \sim 250$ GeV,
this mixing is extremely small, and can be neglected.  Hence, we see
that $h$ behaves essentially like the SM Higgs  (except of course in
interactions with the neutrinos).

It is true that this model requires considerable fine-tuning, in order
to  maintain the hierarchy
$\langle V_{\phi}\rangle / \langle V_{\chi}\rangle \sim 10^{-13}$, 
which is not naturally stable with respect to quantum corrections. 
However, this is no worse than the case in the usual 
non-supersymmetric grand unified theories, and also in the type II
seesaw models for neutrino masses involving Higgs triplets.
It will be interesting to see if the model can be supersymmetrized to
resolve this. In any case, since the scenario suggested here 
has experimental signatures of a strikingly novel kind, we feel
that its consequences are certainly worth exploring, much in the
same way as the phenomenology of various other models have been
explored in recent times.

From Eq.(\ref{phi}), we see that the charged Higgs mainly resides
in the doublet $\phi$, with only a very tiny part, $V_{\phi}/V$ in
$\chi$. Thus the coupling of the charged Higgs with the quarks are
highly suppressed (hence the Chromophobic charged Higgs). However
its coupling with the neutrinos and the corresponding charged
leptons are not suppressed. Thus the charged Higgs will dominantly
decay to the neutrinos and the charged leptons, giving a totally
different signals from the usual generic two Higgs doublet models,
or the MSSM. The Yukawa couplings of the charged Higgs, H to the
leptons and quarks are given by
\bea{chc} L_Y = - \sqrt{2} \frac{m_{\nu}}{V_{\phi}} r_{\chi}
[\overline{l_L} \nu_R H + \overline \nu_L l_R H  + h.c. ] \nonumber \\
+  \sqrt{2} \frac{m_d}{V_{\chi}} r_{\phi}\overline{u_L} d_R H
- \sqrt{2} \frac{m_u}{V_{\chi}} r_{\phi}\overline {d_L} u_R H  + h.c.
\eea
where $r_{\chi}=V_{\chi}/V$ and $r_\phi=V_{\phi}/V$.
The Feynman rules for the interaction of the charged Higgs with the
photon, $Z$ boson and the scalars $\sigma$ and $\rho$ are given by

\begin{center}
\begin{tabular}{|l|l|} \hline
  Fields                                   &  Couplings             \\[2mm]\hline
${A}_{\mu} (p_1)$ $H^+(p_2)$ $H^-(p_3)$    & $e\big(p_3-p_2\big)^\mu$\\[2mm]
${Z}_{\mu} (p_1)$ $H^+(p_2)$ $H^-(p_3)$    & $\frac{(1-2s_W^2)e}{2s_W c_W}\big(p_3-p_2\big)^\mu$\\[2mm]
${H}^+(p_1)$ $\sigma (p_2)$ $W^-_\mu(p_3)$ & $\frac{er_\chi}{2s_W}\big(p_2-p_1\big)^\mu
$\\[2mm]
${H}^-(p_1)$ $\rho (p_2)$ $W^+_\mu(p_3)$   & $\frac{ie}{2s_W}\big(p_2-p_1\big)^\mu $\\[2mm]\hline
\end{tabular}
\end{center}

The following important features of this model become apparent
from the above description:
\begin{itemize}
\item  The charged Higgs $H^\pm$ has
 practically no coupling to a pair of quarks (that is to say, these are
`chromophobic' scalars).

\item While $h,\rho$ and $H^\pm$ have masses in the electroweak
scale, $\sigma$ is an extremely light physical state whose mass is
controlled by the  vev $V_\phi$. $\sigma$ has interesting
cosmological implications which was discussed in Ref.\cite{gabriel:2006}.

\item The coupling of the charged scalar physical states $H^{\pm}$
to a lepton and the corresponding neutrino is large, proportional to
the mass of the neutrino in that family divided by a tiny vev of the
order of the corresponding neutrino mass.

\item The main decays of  $H^{\pm}$ are
$H^{\pm}\longrightarrow \ell\nu_\ell$($\ell = e/\mu/\tau$),
$H^{\pm}\longrightarrow \rho W^{\pm}$, and $H^{\pm} \longrightarrow
\sigma W^{\pm}$.
\end{itemize}

It is also to be noted that the absence of interaction with quarks
makes the charged Higgs in this scenario free from all constraints
arising from rare processes such as $b \longrightarrow s\gamma$.
Thus its mass can be anything above the limit from the search for
pair-production at the Large Electron Positron (LEP) collider.

\section{LHC signals for the charged Higgs}

\begin{figure}[t]
\begin{center}
\includegraphics[height=3.5in,width=3.5in]{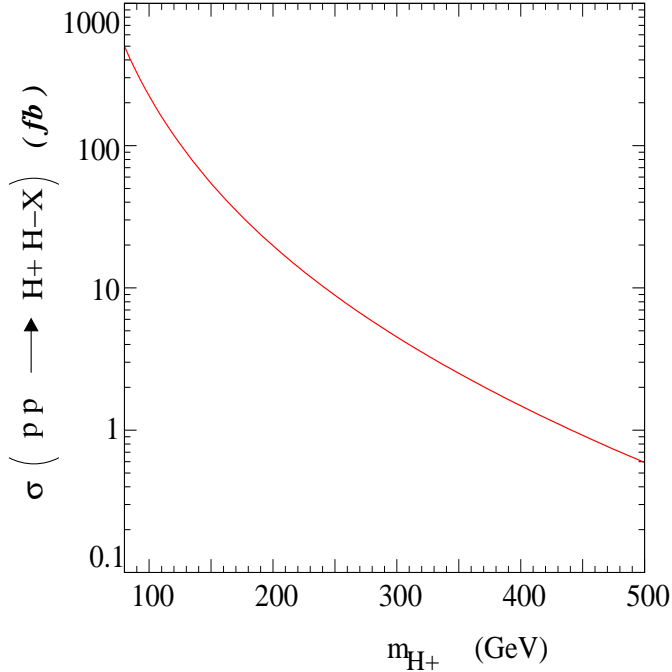}
\caption{\it Cross section for the pair production of 
charged Higgs at LHC as
a function of the charged Higgs mass. We have used the leading order parton
density functions of CTEQ6L \cite{cteq} for the analysis.}
\label{fig:pairprod}
\end{center}
\end{figure}

The chromophobic property of $H^\pm$ makes its search channels at the
LHC quite different from the usual ones. The usual search strategies for
charged Higgs at hadron colliders rely on its associated production
with top quarks or from top quark decays. These production channels are denied
in our case due to the chromophobic property of the charged Higgs. 
First of all, its production cannot take place through the process 
$bg \longrightarrow t H^-$ \cite{bgthplus} or through top decays, $t\to b H^+$. 
One has to depend on electroweak processes leading to its pair-production.
The pair productions of the charged Higgs in our model is via
Drell-Yan process with the exchange of the photon and the Z boson in
the  s-channel. One could also have the charged Higgs boson pair produced at
LHC through scattering of two electroweak gauge bosons via 
$qq \to qqV^*V^* \to qq H^+H^-$ where $V=\gamma,Z,W^\pm$ \cite{moretti}. 
However we note that this cross section is suppressed compared to the Drell-Yan
process. The production cross section for the charged Higgs pair through the 
Drell-Yan channel at the LHC energy are shown in Fig.\ref{fig:pairprod}.

We can also produce the charged Higgs singly at LHC through the associated 
production channel $qq' \to \rho H^\pm, \sigma H^\pm$. Both the production
modes would lead to a single charged Higgs in the final state in association 
with large missing transverse energy as $\rho$ decays to a pair of neutrinos 
with 100\% branching ratio while $\sigma$ passes through the detectors 
undetected \cite{gabriel:2006}. For the single-$H^\pm$ production channel, 
although the rates are of magnitude comparable to that of pair-production, 
the single-W background turns out to be overwhelmingly large, the reason
being the substantial branching ratio of the W to either an
electron or a muon. Thus the search for the charged Higgs in 
this scenario is best carried out via pair-production.

\begin{figure}[t]
\begin{center}
\includegraphics[height=3.2in,width=3.2in]{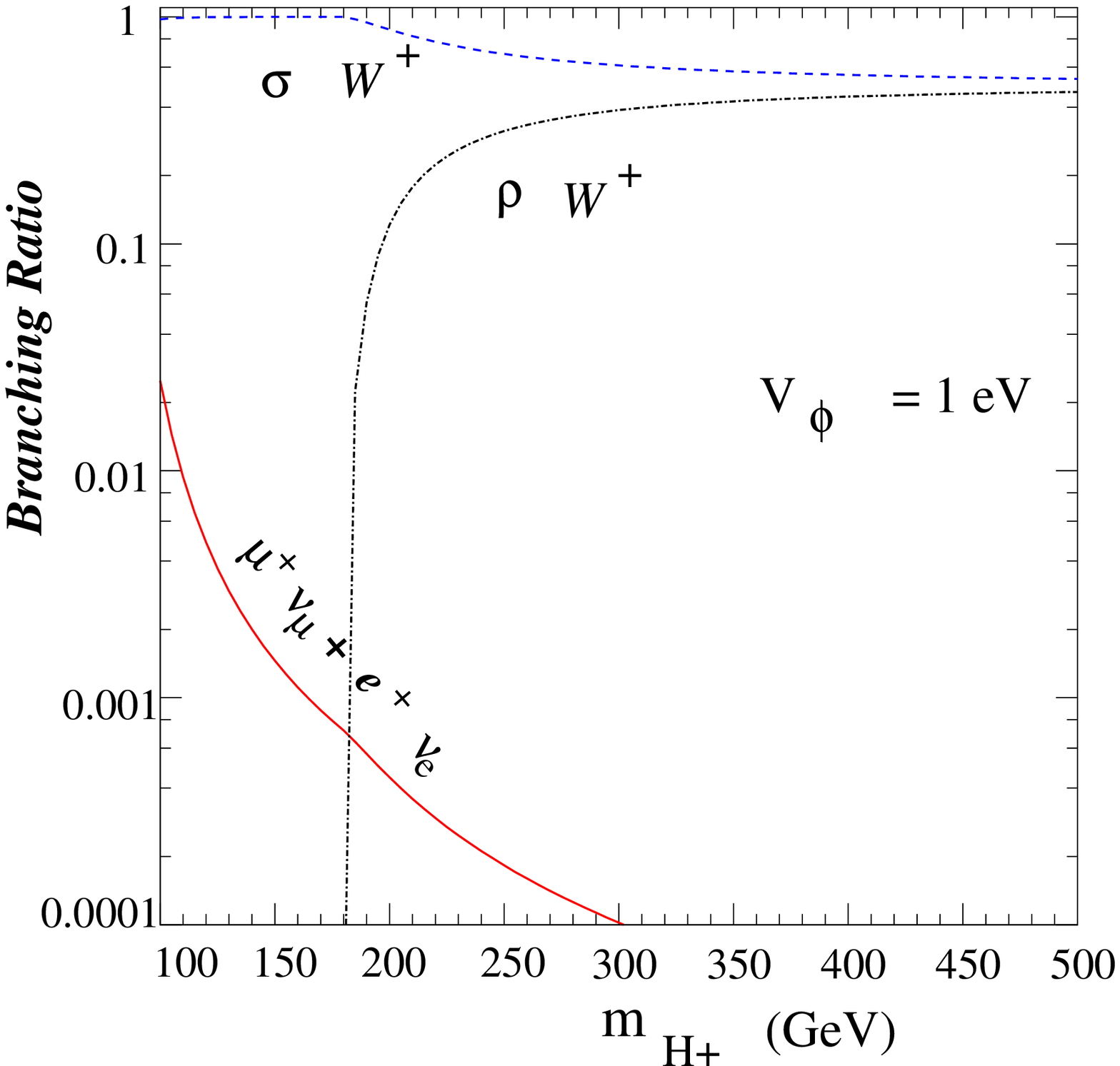}
\includegraphics[height=3.2in,width=3.2in]{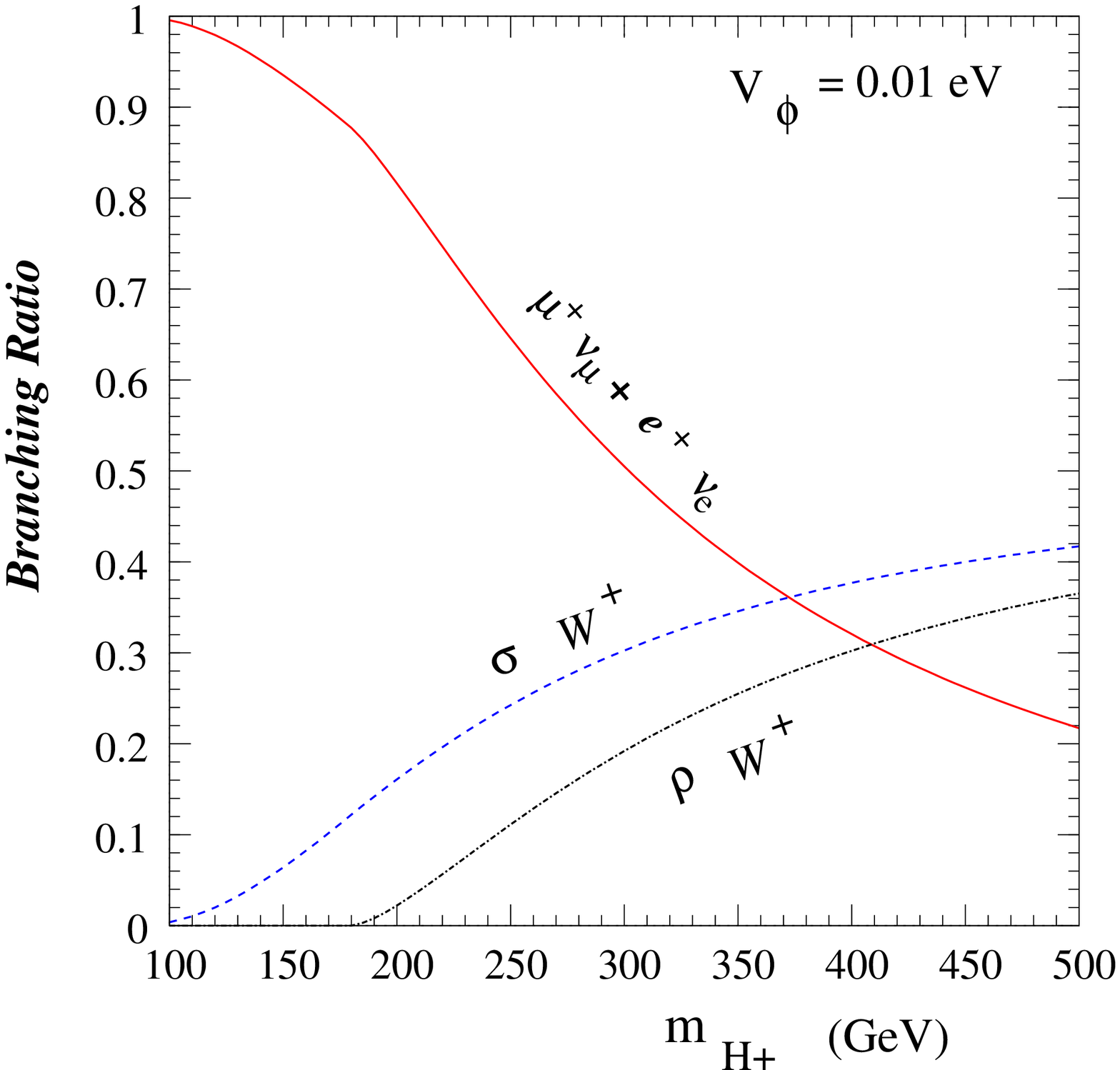}
\caption{\it We plot the branching ratio for the decay of charged Higgs boson
as a function of its mass. We also show, how the value of the vev for the
second doublet affects the branching ratios.}
\label{fig:Hdk}
\end{center}
\end{figure}

Decay branching ratios of the charged Higgs are determined by a
competition between the neutrino - charged lepton and the $\sigma W$ and
$\rho W$ final states, the respective branching ratios being decided
by the vev $V_\phi$ which in turn determines neutrino masses.  Plots
of the branching ratios are shown in Fig.\ref{fig:Hdk}, where one finds that
the fermionic decay modes are more favoured for (a) low charged
Higgs masses , and (b) relatively smaller values of $V_\phi$. The
dominant fermionic decay mode is in the channel $\tau\nu_\tau$ in
the case of normal hierarchy of neutrino masses.  However, in the
inverted hierarchy scenario the dominant fermionic decay modes are
to $\mu\nu_\mu$,$e\nu_e$ . Therefore, in case the IH scenario is
preferred by nature, this model predicts the rather striking
signature for the charged Higgs, namely,

$pp \longrightarrow H^+ H^- \longrightarrow \ell^+ \ell'^- + {E\slash_T}$ \\
(with $\ell,\ell' = e/\mu$).

With the above final states in mind, we calculate the event rates
for the signal and compare them with the SM background.
We have chosen a sample (benchmark) point in the parameter space of the 
model for our analysis. The choice for the free parameters of the theory are
\begin{itemize}
\item $\l_1=0.12,~\l_2=1.0, ~\l_3 = 2.0$
\item $M_\rho=100$ GeV, $M_\sigma\sim V_\phi=10^{-2}$ eV and $M_h=120$ GeV.
\item $\l_5=\frac{M_\rho^2}{V^2}$ and $\l_4=\frac{2m_{H^\pm}^2}{V^2}-\l_5$.
\end{itemize}
It is worth pointing out that the charged Higgs production rate is not affected
for other choices of the $\l_i's$ allowed by the model. However, as
Fig.\ref{fig:Hdk} reveals, the choice of $V_\phi$ plays a 
crucial role in the decay properties of the 
charged Higgs. Also, the mass of the pseudoscalar $\rho$
influences the branching ratios for $m_{H^+}\le m_W+M_\rho$ to
some extent.

The SM background mainly comes from the process
$pp \to W^+W^-$ and $pp \to ZZ$. In the first case the $W$-bosons decay to
$e/\mu$ and a neutrino, while in the second, one of the $Z$
decays into neutrino and the other goes to an electron/muon pair.
The second channel can be suppressed by removing the $Z$-peak in the 
invariant mass distribution of the charged lepton pair.
In addition, a strong $E\slash_T$ cut (which retains an appreciable
fraction of the signal due to the larger mass of the charged Higgs)
helps in reducing the SM background.
It is important to note that, although the $W$-pair production
cross section is quite large at LHC ($\sim 120~pb$)\cite{haywood}, the 
small branching ratio to $\ell\nu_\ell$ reduces the effective background rate 
as compared to the signal, for which the branching fraction is large when 
the $V_\phi$ is small (but approximately in the right range to yield proper 
neutrino masses with Yukawa couplings $O(1)$), and the charged Higgs mass is
$\le$ 300 GeV. 

However with additional jets coming from 
initial and final state radiations off the colliding partons, one expects the 
signal to be accompanied with jets. This leads to another major source for the 
SM background coming from the $t\bar{t}$ production\footnote{We use the 
available NNLO corrected cross-section $\sim 890~pb$ \cite{nnlottbar}.}. 
At first thought, this should be reducible by tagging the $b$-jets with 
large $E_T$ in the final state. Assuming an efficiency of 60\% for a 
single $b$-jet identification, this should eliminate 84\% of the background 
coming from the $t\bar{t}$ production. 
However, this does not prove sufficient to completely reduce this background. 
We note that this background can be effectively suppressed without 
losing much of the signal if we put a selection criterion on the maximum 
number of jets associated with the signal. Keeping this in mind we perform 
our analysis.
\begin{figure}[htb]
\begin{center}
\includegraphics[height=3.5in,width=3.5in]{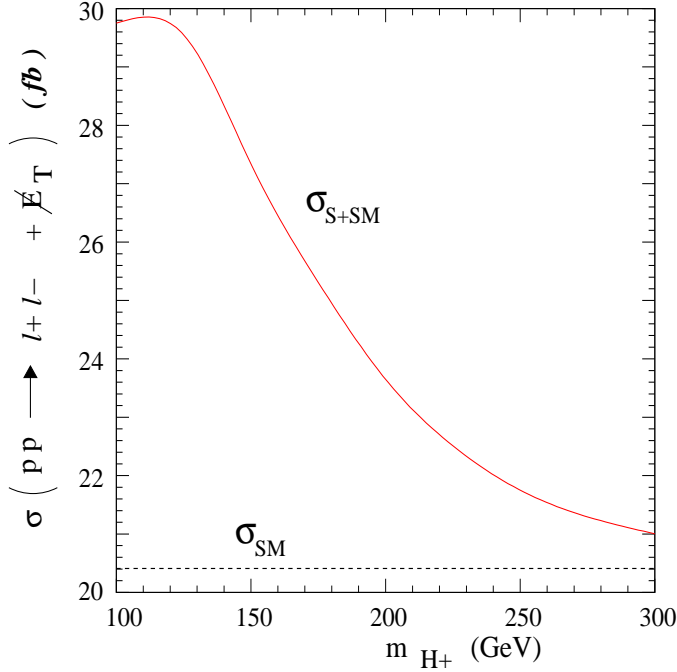}
\caption{\it We show the total cross section for the process
$pp\to \ell^+ \ell^- + E\slash_T$ as a function of the charged Higgs mass
at the LHC, where $\ell = e,\mu$.
The SM cross section is also shown in broken lines.}
\label{fig:twotau}
\end{center}
\end{figure}

The SM background event generation has been done using 
PYTHIA 6.410 \cite{pythia}. The signal events have been generated using the 
CalcHEP 2.4.5 \cite{calchep} package and then interfaced with PYTHIA. To define 
the associated jets we employ the jet cone algorithm implemented in PYTHIA
through the subroutine PYCELL. The minimum $E_T$ threshold for a cell to be 
considered as a jet initiator has been chosen as 2 GeV, while we assumed the 
minimum summed $E_T$ of the jet (consisting of all cells within the cone of 
radius $R$ in the ($\eta,\phi$) plane) to be accepted as 
a jet to be 20 GeV. The jet conical width is 
$\Delta R_{jj}=\sqrt{\eta^2_{jj}+\phi^2_{jj}}\ge 0.7$. while $\eta$ coverage 
range for jets is taken to be $|\eta|\le 3.0$. Using the above clustering 
algorithm we find that if we restrict ourselves to $N_{jets} \le 2$ 
(where  $N_{jets}$ is the total number of jets) and using
the $b$-jet identification efficiency, the background from the $t\bar{t}$ 
production is reduced to about $2-3~fb$ after implementing the selection cuts 
listed below. The dominant background still remains the one arising from 
the $W^+W^-$ production.

Based on the above observations and restricting ourselves to $N_{jets} \le 2$, 
we have imposed the following cuts on our final state events:
\begin{itemize}
\item The transverse momentum of the charged lepton should respect a minimum 
cut $p_T^\ell > 25$ GeV.
\item The charged leptons should be in the rapidity interval
$|\eta_\ell| < 2.5$.
\item A missing transverse energy(momentum) cut given by
$E\slash_T > 100$ GeV.
\item The $e/\mu$
should be well separated in space to be resolved, thus justifying
$\Delta R_{\ell\ell} \ge 0.4$ where
$\Delta R = \sqrt{(\Delta \eta)^2 + (\Delta \phi)^2}$.
\item $M^{inv}_{\ell\ell} > 100$ GeV.
\end{itemize}
In Fig.\ref{fig:twotau} we show a plot of the signal rate against 
the charged Higgs mass. The backgrounds are represented by the 
horizontal line. We see from Fig.\ref{fig:twotau} that 
for $m_H < 150$ GeV, the signal to background ratio is greater than 1/3, while
it falls to 1/6 for a 200 GeV charged Higgs. For $m_H \simeq 300$ GeV 
this ratio falls to less than 1/30.
It is clear from the figure that, although the
background is sizable, such statistical significance as to set the
signal apart can be achieved with sufficient integrated luminosity
\begin{figure}[htb]
\begin{center}
\includegraphics[height=3.2in,width=3.2in]{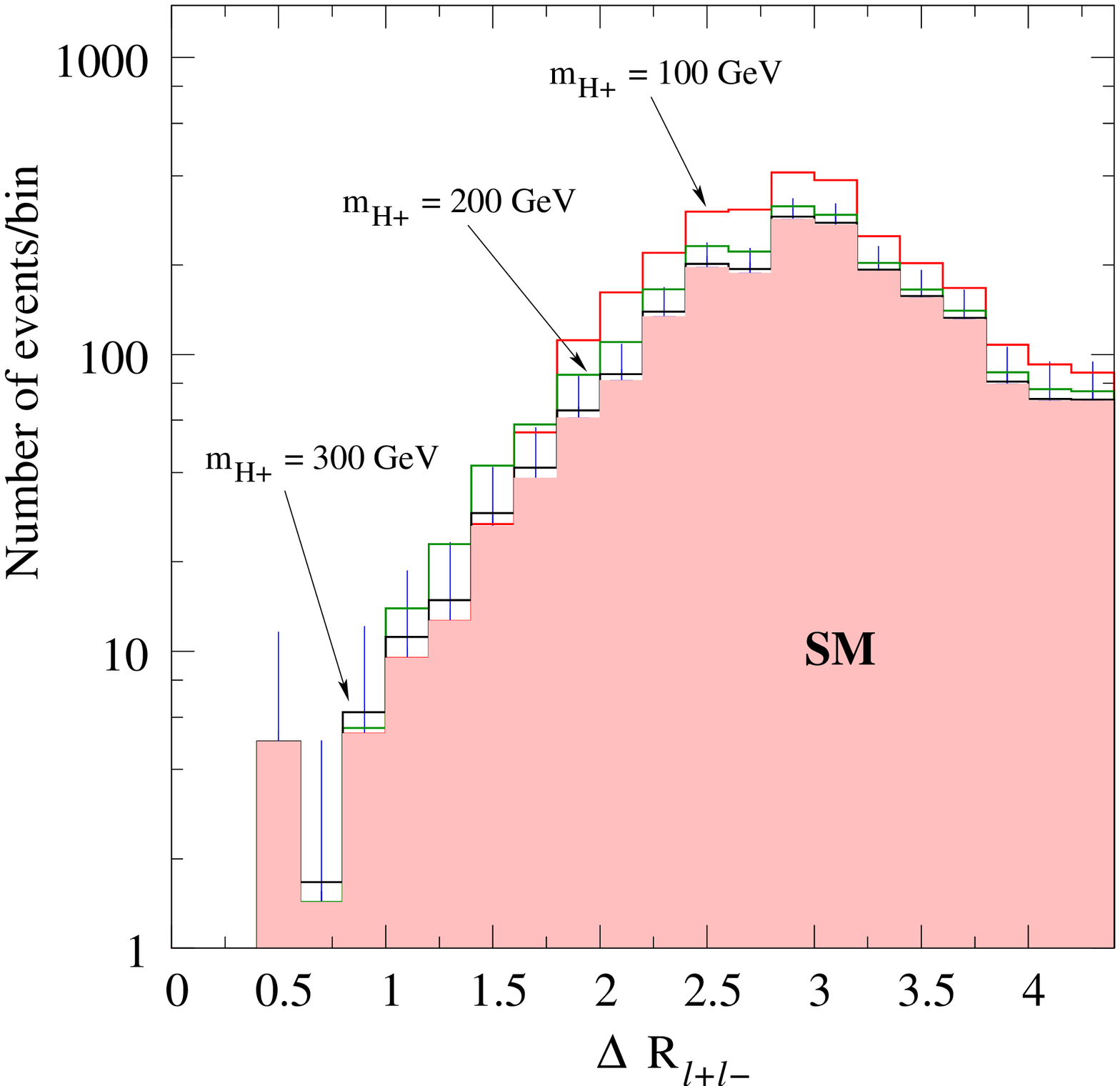}
\includegraphics[height=3.2in,width=3.2in]{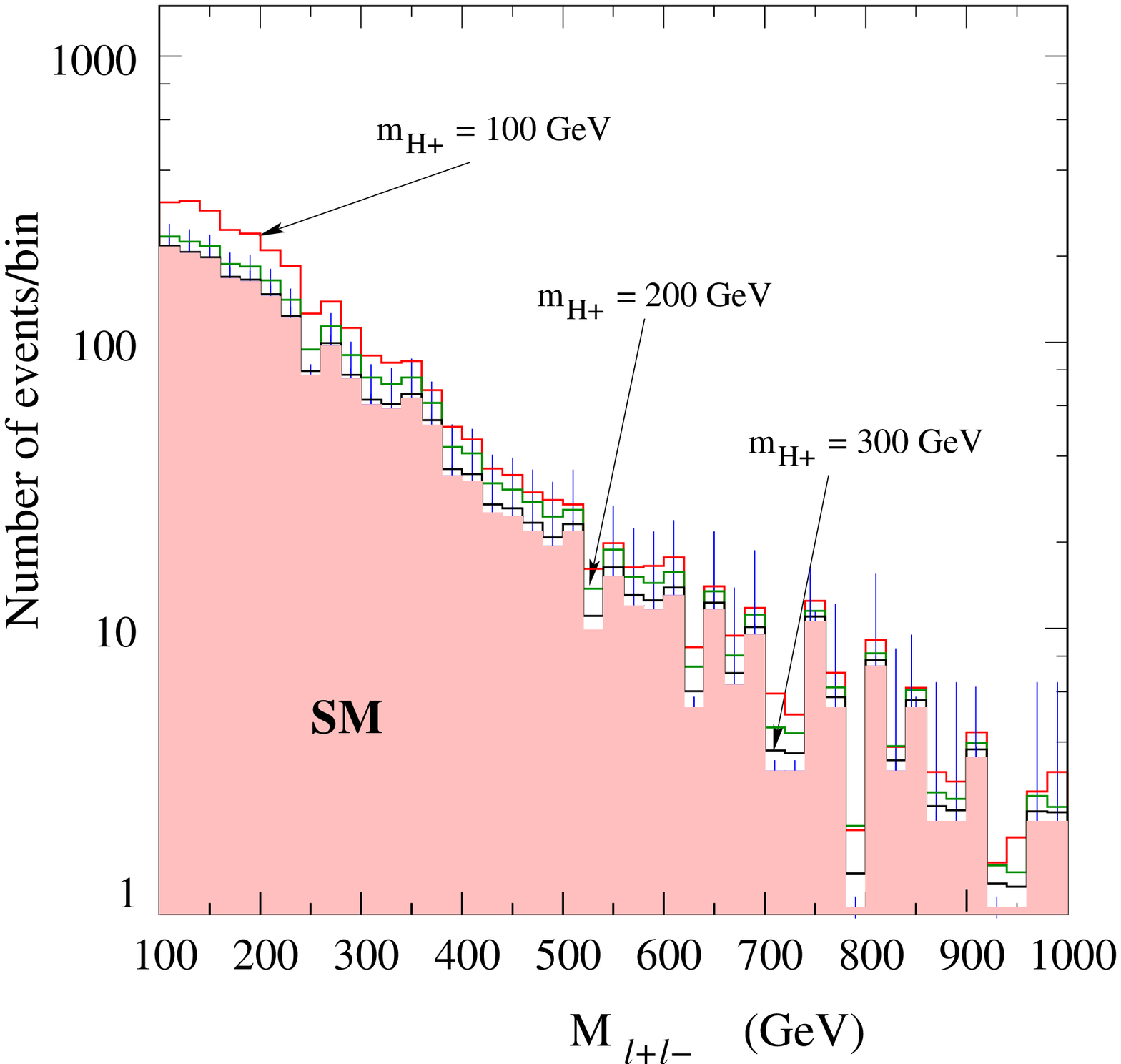}
\caption{\it We show the binwise distributions in the $\Delta R_{\ell^+\ell^-}$
 and the invariant mass $M_{\ell^+\ell^-}$ for the $\ell^+\ell^- + E\slash_T$
signal for three different choice of mass of the charged Higgs. The SM
expectation is also shown in the shaded region. The vertical blue bands over
the SM distribution represent the $3\sigma$ statistical fluctuations in the SM
background. The integrated luminosity is taken as $\mathcal{L}=100~fb^{-1}$.}
\label{fig:twotau1}
\end{center}
\end{figure}
at the LHC. For an integrated luminosity of 10 $fb^{-1}$, one 
has a 5$\sigma$ significance for $m_{H^\pm}\le$ 140 GeV while, 
if for example, one has $\int {\cal L}dt~=~100~fb^{-1}$ of
luminosity, then one has a $\sim$5$\sigma$ significance for $m_{H^\pm}\le$
220 GeV. The search limit goes up to about 250 GeV with the same
statistical significance for $\int {\cal L}dt~=~300~fb^{-1}$. Since
the charged Higgs mass has little constraint on it other than that
from Drell-Yan pair-production at the LEP, the above result is quite
encouraging, as one is probing a substantial part of the parameter
space of a chromophobic charged Higgs answering to an inverted
hierarchy of neutrino masses.

\begin{figure}[htb]
\begin{center}
\includegraphics[height=3.2in,width=3.2in]{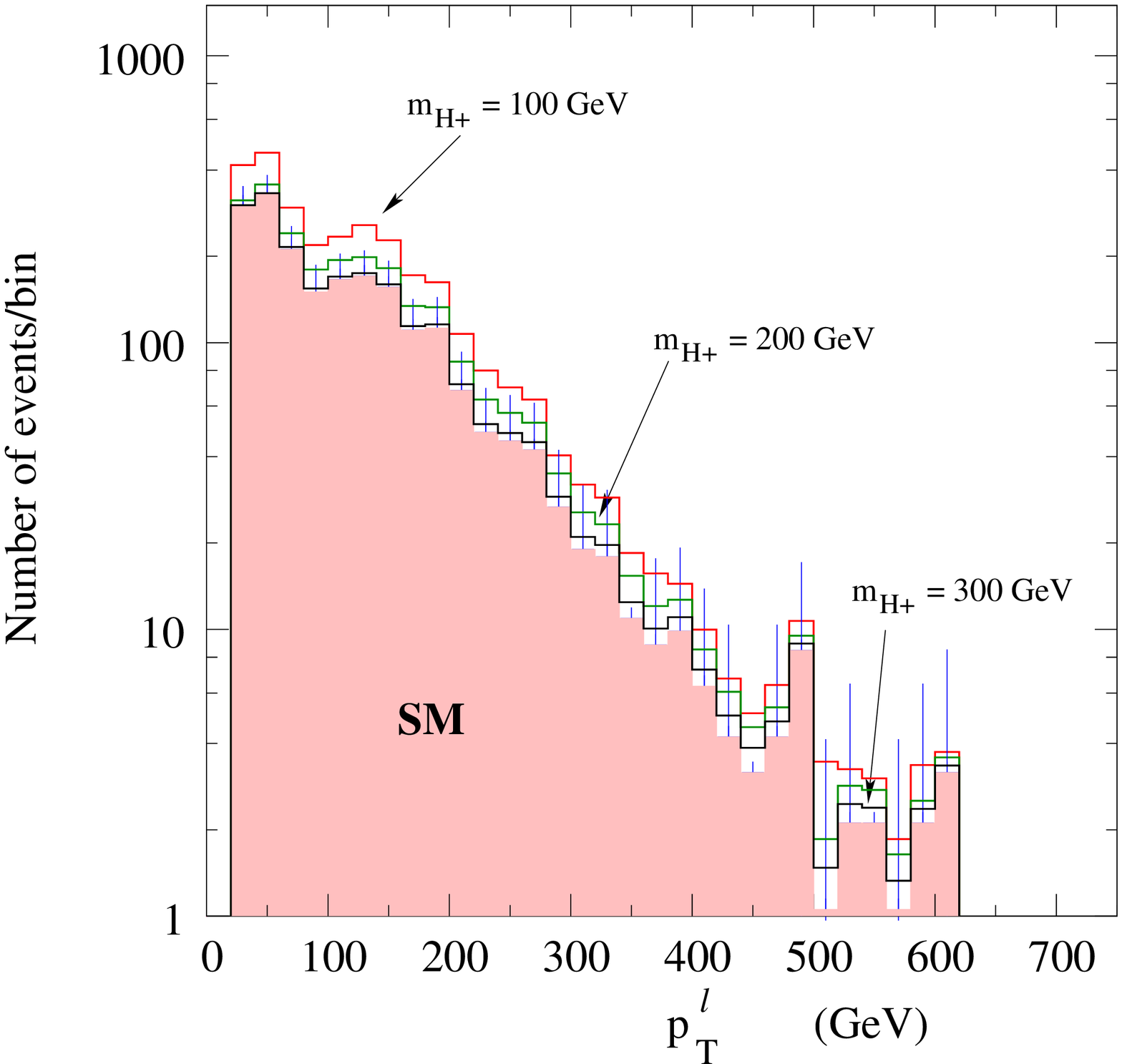}
\includegraphics[height=3.2in,width=3.2in]{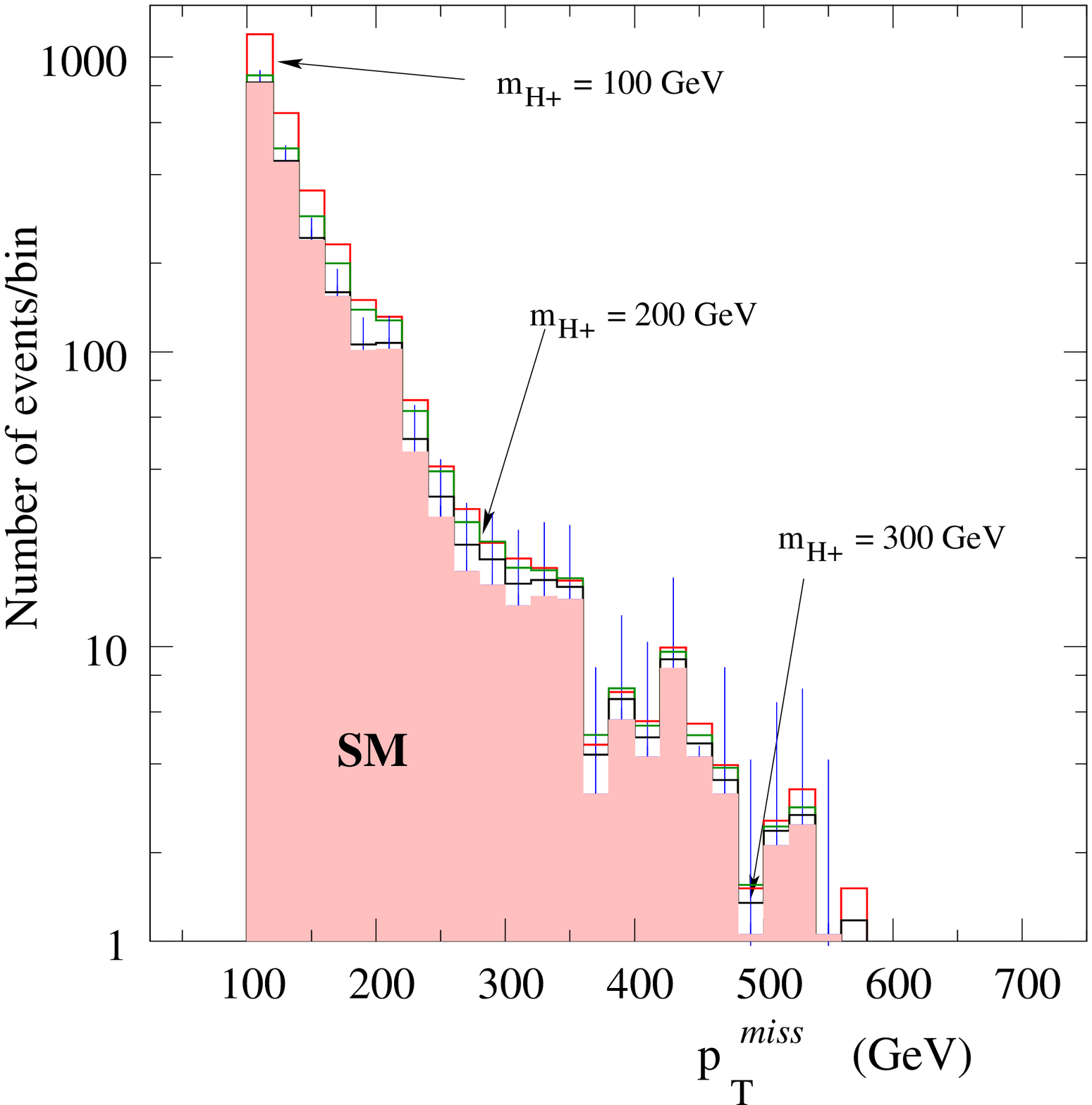}
\caption{\it We show the binwise distributions in the $p_T^{\ell}$ 
and the $E\slash_T$ for the $\ell^+\ell^- + E\slash_T$ signal for 
three different choice of mass of the charged Higgs. 
All conventions are the same as in Fig.\ref{fig:twotau1}.
The integrated luminosity is taken as $\mathcal{L}=100~fb^{-1}$.}
\label{fig:twotau2}
\end{center}
\end{figure}
We also present some kinematic plots of the signal in
Fig.\ref{fig:twotau1} and Fig.\ref{fig:twotau2} and compare it with
the SM background. The distributions show how the signal stands out
better with larger $m_{H^\pm}$ so long as the production rate due to
higher mass is not forbiddingly low.

In the case of a normal hierarchy of neutrino masses, the dominant
coupling  of $H^\pm$ is to a $\tau-\nu_\tau$ pair. The corresponding
signal is $\tau^+ \tau^- + E\slash_T$, for which the rates without
any cuts is same as that for the $e/\mu + E\slash_T$ final state,
since the branching ratio for $H^\pm \longrightarrow \tau\nu_\tau$
in NH is the same as that for  $H^\pm \longrightarrow (e \nu_e +
\mu\nu_\mu)$ in IH. The backgrounds, on the other hand, are 
reduced by a factor of four due to the smaller branching ratio of
each W decaying into $\tau\nu_\tau$ only. Thus one expects {\it prima
facie} a better search limit for the $H^\pm$ in this case. However,
one has to study the effects of $\tau$-decays and the cuts
on the decay products more carefully. An available option is to
identify $\tau$-polarisation and thus separate the signals from
the W-backgrounds, for which the polarisation is of opposite type.
A detailed quantitative study of this signal pertaining to the
NH case will be reported in a subsequent paper \cite{inprep}. 

\section{Conclusions}

In a model motivated to explain the tiny neutrino masses, we have
discussed a scenario in which the charged Higgs productions and
decays are completely different from the usual two Higgs doublet
models or MSSM. In this model, the dominant charged Higgs pair
productions are via the Drell-Yan processes, where its decays are
dominantly to the charged leptons and the corresponding neutrinos.
In the inverted neutrino mass hierarchy scenario, the dominant decay
modes are to the light leptons, electrons and muons. Such signals
can be detected at the LHC for a charged Higgs mass upto few hundred
GeV, and will provide a clue to the pattern of the neutrino masses.

\vspace*{0.4in}
\textbf{Acknowledgement}

We thank C. Csaki, A. Deroeck, D.K. Ghosh, K. Huitu, T. Han and S. Roy 
for useful discussions. SN would like to thank the CERN Theory Division for 
warm hospitality and support during his sabbatical there where part of this 
work was done. BM and SN would like to thank the warm hospitality and 
support of KITP, Santa Barbara, and the the organizers of the Workshop
"Physics of the Large Hadron Collider" during their participation
there when this work was completed. The work of SG and SN was
supported in part by the US Department of Energy, Grant Numbers
DE-FG02-04ER41306 and DE-FG02-ER46140. The work of BM was partially
supported by funding available from the Department of Atomic Energy,
Government of India, for the Regional Center for Accelerator-based
particle Physics, Harish-Chandra Research Institute. SKR gratefully
acknowledges support from the Academy of Finland (Project
No.115032). This research was supported in part by the National
Science Foundation under Grant Number PHY05-51164.

\end{document}